\documentclass[11pt,a4paper]{llncs}
\usepackage{amsmath}
\usepackage{amssymb}
\usepackage{graphicx}
\usepackage{marvosym}
\usepackage{url}
\usepackage{fancyhdr}

\usepackage{geometry}
\newcommand{\keywords}[1]{\par\addvspace\baselineskip
\noindent\keywordname\enspace\ignorespaces#1}

\fancypagestyle{firstpage}{\fancyhf{}
}

\begin{document}


\title{\LARGE{A Multi-Agent Retrieval-Augmented Framework for Work-in-Progress Prediction}}


%
%
\author{\large{Yousef Mehrdad Bibalan\textsuperscript{1}  \and Behrouz Far\textsuperscript{1} \and \\ Mohammad Moshirpour\textsuperscript{2} \and Bahareh Ghiyasian\textsuperscript{3}}}
\institute{\large{\textsuperscript{1}University of Calgary, Canada,\\ \textsuperscript{2}University of California, Irvine, USA, \\\textsuperscript{3}Google, USA}}

%


%
%


\maketitle

\thispagestyle{firstpage}

\begin{abstract}
Work-in-Progress (WiP) prediction is critical for predictive process monitoring, enabling accurate anticipation of workload fluctuations and optimized operational planning. This paper proposes a retrieval-augmented, multi-agent framework that combines retrieval-augmented generation (RAG) and collaborative multi-agent reasoning for WiP prediction. The narrative generation component transforms structured event logs into semantically rich natural language stories, which are embedded into a semantic vector-based process memory to facilitate dynamic retrieval of historical context during inference. The framework includes predictor agents that independently leverage retrieved historical contexts and a decision-making assistant agent that extracts high-level descriptive signals from recent events. A fusion agent then synthesizes predictions using ReAct-style reasoning over agent outputs and retrieved narratives. We evaluate our framework on two real-world benchmark datasets. Results show that the proposed retrieval-augmented multi-agent approach achieves competitive prediction accuracy, obtaining a Mean Absolute Percentage Error (MAPE) of 1.50\% on one dataset, and surpassing Temporal Convolutional Networks (TCN), Long Short-Term Memory (LSTM), and persistence baselines. The results highlight improved robustness, demonstrating the effectiveness of integrating retrieval mechanisms and multi-agent reasoning in WiP prediction.

\end{abstract}

\section{Introduction}
Predictive process monitoring is essential in modern management because it helps forecast workload changes and supports effective resource planning \cite{ceravolo2024predictive}. A key task in predictive process monitoring is work-in-progress (WiP) prediction—estimating the number of active tasks at any given moment—which improves staffing, capacity planning, and overall operational efficiency. Traditional predictive process monitoring techniques employ deep learning models such as recurrent neural networks, long short-term memory networks, and transformers to capture sequential dependencies in event logs \cite{rama2021deep,wang2023mitfm,tax2017predictive,nguyen2021time}. More recently, narrative encoding has been introduced to enrich these models by transforming raw event traces into structured textual stories. Methods such as LUPIN and SNAP fine-tune pre-trained large language models (LLMs) to leverage this natural-language representation, yielding improved predictive accuracy \cite{pasquadibisceglie2024lupin,oved2025snap}. Although narrative approaches enhance context modeling, they remain purely generative and are typically confined to a single contextual perspective.

Retrieval-augmented generation (RAG) offers a compelling way to ground generative predictions in concrete past cases by fetching relevant documents or data fragments that augment the model’s input \cite{cheng2025survey}. RAG has achieved notable success in domains such as open-domain question answering and time-series forecasting, yet it has not been systematically applied to predictive process monitoring \cite{gruver2023large}. Separately, agentic LLM architectures-where multiple specialized agents collaborate under an orchestrator-have been proposed for diagnostic tasks in process mining but focus solely on retrospective analysis and lack retrieval components \cite{berti2024crewai,xuopenrca}. To the best of our knowledge, no existing work combines retrieval-augmented generation with multi-agent reasoning for WiP prediction.

To address these gaps, this paper proposes a retrieval-augmented, multi-agent framework for predicting WiP. Event logs are transformed into temporal and semantic narratives, indexed in a vector-based memory for RAG-based retrieval. Predictor agents, each aligned with a specific narrative view, generate individual predictions in a zero-shot manner, without any task-specific fine-tuning. A decision-making assistant extracts high-level signals, such as momentum and variability, to guide interpretation. A fusion agent then integrates these predictions and insights using ReAct-style reasoning to produce a robust WiP prediction. By integrating retrieval at every stage and orchestrating agent collaboration, our framework grounds its outputs in both agent-generated forecasts and relevant historical cases, ensuring accuracy and contextual sensitivity.

This paper makes three main contributions as follows: 
\begin{itemize}
  \item Transform raw event logs into rich narrative stories that capture multiple temporal and semantic dimensions of process behavior, enabling LLMs to reason over structured, human-readable contexts. 
  
  \item Show that RAG can serve as a dynamic memory-replacing traditional in-memory storage by fetching relevant past cases to inform forecasts and improve transparency. 
  
  \item Develop a zero-shot, multi-agent system in which specialized predictor agents, a decision-making assistant agent, and a fusion agent collaborate-integrating narrative perspectives and retrieved evidence-to produce accurate WiP predictions.

\end{itemize}

\section{Related Work}
Predictive process monitoring aims to forecast the future behaviour of ongoing process instances using historical event logs \cite{turn0academia21,redis2024skill}. Traditional approaches often rely on sequential models such as LSTMs, GRUs, or encoder-decoder networks, which model the process trace prefix and predict the next activity or suffix \cite{evermann2017predictive,rama2021deep,rama2022encoder}. More recent advancements have explored the potential of LLMs for these tasks \cite{berti2024pm}. For example, Xu and Fang proposed LLM4NT, a domain-specific adaptation of the Qwen 2.5B decoder with a cross-attention mechanism, to predict next timestamps using event trace inputs \cite{xu2025next}. Their approach demonstrates strong performance on business process logs, proving that LLMs can generalize beyond natural language domains for predictive tasks. However, most of these LLM-based methods operate in a monolithic fashion and do not incorporate retrieval or agent-based modularity~\cite{xu2025next,beheshti2023processgpt}.

Another prominent development is the generation of semantic narratives from structured event logs. The SNAP framework~\cite{oved2025snap} addresses the limitations of conventional models by constructing natural-language semantic stories from process traces. These stories encode multiple event attributes into a single narrative, enabling small and large language models to learn meaningful representations for next activity prediction. SNAP outperforms 11 state-of-the-art models across six benchmark datasets and shows significant gains in contexts rich in textual or conversational data. However, SNAP relies on a single-shot story template generated by an LLM and lacks a retrieval or agentic layer for dynamic adaptation.

Despite the rapid adoption of RAG in NLP, its application in process mining and predictive monitoring is still emerging and several agentic frameworks exhibit RAG-like behavior on XES-based logs. For instance, Jessen et al.\ proposed a conversational LLM agent that uses prompt grounding via process ontologies and dynamic SQL generation for query refinement~\cite{jessen2023prompt}. Similarly, the CrewAI architecture by Berti et al.\ embeds LLM agents that invoke deterministic tools like PM4Py and SQL to carry out analytic tasks such as conformance checking or root-cause diagnosis~\cite{berti2024crewai}. These architectures capture the essence of RAG, but are primarily analytical and lack integration with predictive modeling workflows.

The use of AI agents in process mining has been explored through orchestrated LLM-powered systems that assign specific responsibilities to different agents. CrewAI, for example, structures agents around process mining tasks, allowing modular decomposition and traceable outputs. Each agent is equipped with specialized prompts and access to deterministic APIs, creating a ``think-act'' architecture suited for explainable automation~\cite{berti2024crewai}. Aratchige and Ilmini provide a comprehensive survey of such agentic architectures and highlight their applicability for transparent decision-making~\cite{aratchige2025survey}.

While foundational work exists on narrative generation, LLM-based suffix prediction, and prompt-grounded analytics, the integration of RAG and multi-agent LLMs for predictive monitoring---especially over structured and semantic process logs---remains an open challenge. The proposed multi-agent RAG+LLM framework aims to bridge this gap by combining retrieval, reasoning, and modular orchestration in a scalable predictive pipeline.

\section{Proposed Approach}
This section describes the architecture and design principles of the proposed framework. We begin by defining the WiP event log structure, then outline how narrative stories and process memory are constructed, followed by the agents’ design and their interactions.

\subsection{WiP Event Log}

An event log, a crucial input for predictive process monitoring (PPM), is defined as a sequence of events $L = \{e_i\}$, where each event is defined as
\[
e_i = (c, a, t, (k_1, v_1), ..., (k_j, v_j))
\]
Here, $c$ denotes the case identifier, $a$ is the activity name, $t$ is the timestamp, and each pair $(k_j, v_j)$ represents an attribute and its corresponding value. This structure captures the temporal and contextual information of a process execution over time. We define a WiP Event Log as an event log that includes WiP events, where each \textit{WiP event} is represented as a 10-dimensional vector:
\[
x = (x_t^{(d_w)}, x_t^{(d_m)}, x_t^{(d_y)}, x_t^{(o)}, x_t^{(h)}, x_t^{(l)}, x_t^{(c)}, x_t^{(n)}, x_t^{(d)}, x_t^{(s)})^T
\]

In this representation, $x_t^{(o)}, x_t^{(h)}, x_t^{(l)}, x_t^{(c)}$ denote the Open, High, Low, and Close values of WiP. The components $x_t^{(d)}, x_t^{(n)}, x_t^{(s)}$ indicate the number of done, new, and started items, respectively, during the time period $t$.

\subsection{Transform WiP Event Log to Stories}
To bridge structured WiP data and natural language reasoning within the framework, we convert each WiP event into a semantically meaningful textual \textit{story} using a LLM. The input to this transformation is a WiP Event Log, and the output is a human-readable story that captures the operational dynamics of the process during a specific time interval. These stories represent the process state in natural language and support retrieval within the framework. Specifically, We generate two types of stories per time unit:
\begin{itemize}
    \item \textbf{Query Story:} A concise, information-rich description used as direct input to the LLM prompt during inference. \\
    \textit{Example:} 
    \textit{``The WiP items opened at 55, reached a high of 70 and a low of 55, before closing at 66, with 10 items completed, 24 new items added, and 21 items started.''} \\
    \item \textbf{Contextual Story:} An extended version used to construct the retrieval corpus, which includes the target value. \\
    \textit{Example:} \textit{``The WiP items opened at 55, reached a high of 70 and a low of 55, before closing at 66, with 10 items completed, 24 new items added, and 21 items started, \underline {while the next WiP was expected to remain at 71.}''}
\end{itemize}

These query and contextual stories function as structured, interpretable knowledge units for retrieval and inference, supporting robust multi-scale forecasting in the framework.

\begin{figure}[!t]  
    \centerline{\includegraphics[width=0.99\columnwidth]{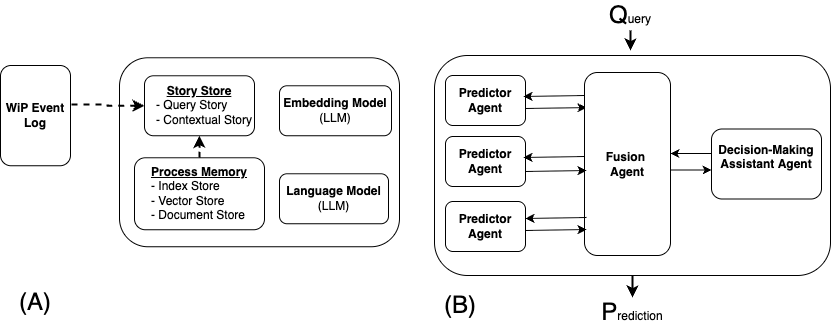}}
    \caption{
    \textbf{Architecture Overview.} 
    (A) Narrative generation and memory construction pipeline: WiP event logs are transformed into query and contextual stories via LLMs and stored in a semantic vector-based process memory. 
    (B) Multi-agent prediction framework, where multiple predictor agents and a decision-making assistant feed into a fusion agent that generates the final WiP prediction.
    }
    \label{framework-architecture}
\end{figure}

Figure~\ref{framework-architecture}(A) provides a high-level view of this narrative generation pipeline. As shown, raw WiP event logs are transformed into query and contextual stories through an LLM-based abstraction process. These stories are then prepared for downstream use in the broader prediction architecture, which will be described in the following sections.


\subsection{Retrieval-Augmented Generation as Process Memory} \label{subsec:process-memory}
To enable RAG-based prediction, we incorporate a \textit{Process Memory Module} that functions as a semantic retrieval system, storing and retrieving contextual knowledge from previously observed stories. Each story is embedded using an embedding model. Each document in memory corresponds to a story and is stored in a vector store. During inference, a new query story is used to retrieve the top-n most relevant contextual stories based on cosine similarity. This process memory supports \textit{Contextual Grounding} by injecting real, comparable past examples into the prediction prompt, and \textit{Temporal Generalization} by facilitating pattern recognition over multiple time scales. This memory-augmented structure enables effective reasoning without task-specific fine-tuning while maintaining high contextual relevance and semantic fidelity. By dynamically incorporating similar historical process states, the memory-based retrieval mechanism supports more accurate and temporally grounded predictions within the RAG framework. 

\subsection{Agents}
Our proposed framework employs a team of specialized agents, each assigned a distinct role in the prediction process. Each agent is equipped with its own process memory, and these agents operate collaboratively to interpret historical context, assist in decision-making, and synthesize predictions using information retrieved from their respective memories. The remainder of this section outlines the roles and interactions of the three types of agents in the framework. \newline

\noindent\textbf{Predictor Agent}\newline
The prediction component is built around a modular ensemble of Predictor Agents, each specialized for a different temporal view of the process history. Forecasting is decomposed into agents that independently retrieve and reason over historical data slices. Each agent accesses a process memory module for relevant context and operates as a callable tool within the decision strategy. This design enables the system to capture short-term fluctuations, periodic patterns, and long-term trends simultaneously, improving accuracy. The framework is scalable and extensible-new agents can be added to support emerging patterns or domain-specific needs. Together, these agents form the core of a robust, modular, and data-grounded WiP predicting architecture. \newline

\noindent\textbf{Decision-Making Assistant Agents}\newline
Complementing the Predictor Agents, Decision-Making Assistant Agents extract high-level patterns and signals-such as trends, anomalies, or workload shifts-from recent process behavior. While they do not produce numeric forecasts, they provide contextual insights that help guide the prediction strategy. Each agent analyzes recent data using lightweight statistical or heuristic methods. Their qualitative outputs, such as trend direction or volatility, inform how the final decision maker (the Fusion agent) weighs and combines predictions. By interpreting evolving process dynamics, assistant agents enhance the system’s robustness and adaptability, allowing it to respond to concept drift and structural changes. \newline

\noindent\textbf{Fusion Agent}\newline
At the center of the architecture is the Fusion Agent, responsible for coordinating the Predictor and Assistant agents for final prediction. Unlike specialized agents focused on individual tasks, this agent performs higher-level reasoning-evaluating retrieved evidence, incorporating decision-support insights from assistant agents, and selecting or combining predictions. Using a ReAct-style prompting strategy, it interacts iteratively with tools, queries the process memory, and reconciles agent outputs via heuristics or consensus logic. This flexible workflow enables context-sensitive forecasting that adapts to evolving process conditions. The agent abstracts multi-agent complexity and provides a unified prediction interface. Its modular design supports extensibility, allowing the integration of new agents. \newline

Figure~\ref{framework-architecture}(B) illustrates how agent outputs flow into the fusion agent to produce the final prediction, completing the multi-agent reasoning layer of the framework.

\section{Experimental Results}
This section presents the experimental setup used to evaluate the proposed framework. We detail the datasets, benchmarks, and story-generation procedure, describe how the process memory and agents are launched, and finally report and analyze the obtained results.

\subsection{Datasets and Benchmarks}
We evaluate our framework on two widely used real-world event logs from the 4TU Center for Research Data: BPIC13 Incidents (783 days of incident-management records) and Helpdesk (1,452 days of support-ticket handling) \cite{bibalan2025maml,bibalan2024tcn}. For baselines, we adopt the three best-performing Temporal Convolutional Network (TCN) configurations from \cite{bibalan2024tcn}, denoted as TCN\#1–\#3, which vary in activation functions and optimizers and achieved the lowest MAPE in prior work. We also include two classical baselines: LSTM and Persistence (Yesterday’s WiP), both reported in \cite{bibalan2024tcn}. This set of benchmarks supports a robust comparison against our multi-agent framework.


\subsection{Building Story Storage on Benchmark Datasets}
Each WiP event is transformed into a semantically rich \textit{story} using GPT-3.5-Turbo, capturing variation across three temporal granularities:
\begin{itemize}
    \item \textit{Daily Stories}: Capture fine-grained, day-level fluctuations in WiP dynamics. These stories describe each day's activity in isolation and do not include any explicit temporal information. 

    \item \textit{Weekday-Aware Stories}: Encode cyclical behavior by incorporating the weekday context (e.g., ``on Monday''), enabling the model to learn patterns tied to recurring time-based phenomena. 

    \item \textit{Windowed Stories}: Aggregate daily data over a rolling window (set to seven in our experiments, but configurable), providing a higher-level view. 
\end{itemize}

These story types enable multi-resolution temporal reasoning, capturing short-term patterns (daily), cyclical behaviours (weekday-aware), and longer-term trends (windowed), each can influence future workload differently. By supporting diversity, the architecture integrates complementary insights and improves overall forecast robustness and adaptability. For each WiP event, we generate two story variants: a \textit{Query Story} and a \textit{Contextual Story} (defined in Section~\ref{subsec:process-memory}). The Query Story is used during inference to construct the input prompt, while the Contextual Story is indexed in the process memory for retrieval. 

\subsection{Enabling Process Memory for Story Storage}
The narrative stories are embedded using the BAAI/bge-base-en-v1.5 model and indexed via the LlamaIndex framework \cite{Liu_LlamaIndex_2022}, enabling efficient dense retrieval over semantically encoded text. Each story is stored as a separate document in a semantic vector index. At inference, the current event is converted into a query story, which retrieves the top five most similar contextual stories based on cosine similarity. These retrieved narratives are incorporated into the input prompt for the predictor agent, grounding the forecast in relevant historical context. To preserve causality, only stories with timestamps prior to the target event are available during retrieval, forming a temporally bounded memory aligned with real-world forecasting constraints.

Although the current setup retrieves all prior stories, the architecture supports configurable strategies, such as limiting to recent weeks or filtering based on semantic similarity or predictive utility. These options can enhance retrieval precision and efficiency, particularly in dynamic or resource-limited environments.

\subsection{Launch Agents}


\textbf{Predictor Agents}\newline
To support prediction across multiple temporal granularities, we design three specialized predictor agents. The \textit{DailyMemoryAgent} captures short-term day-to-day variations, the \textit{WeekdayAwareAgent} models recurring weekday cycles, and the \textit{WindowedAgent} abstracts longer-term trends through using rolling seven-day aggregates. Each agent follows a four-step workflow:

\begin{enumerate}
\item Query construction: Convert the current WiP event into a query story.
\item Context retrieval: Retrieve the top five semantically closest stories from the process memory.
\item Prompt assembly: Build a structured prompt that integrates the query story, task instructions, and retrieved examples.
\item LLM Inference: Send the prompt to the language model and return the predicted WiP.
\end{enumerate}

The inference is performed using OpenAI O3-mini, though the framework allows substitution with any compatible LLM. This modular design enables reasoning across diverse temporal patterns and supports the addition of new agents. It also improves transparency, reusability, and robustness in forecasting WiP under dynamic process conditions.\newline

 \noindent\textbf{Decision-Making Assistant Agent} \newline 
 The system includes a dedicated Decision-Making Assistant Agent, called the \textit{Trend Analyst}, which performs lightweight trend analysis over a rolling 7-day window using a simple moving average. It evaluates the directional change in smoothed values to classify recent workload momentum. Based on the observed trend, the agent outputs a short textual description such as “WiP has been increasing significantly” or “WiP has been relatively stable.” These insights are passed to the Fusion Agent to guide prediction weighting, enabling the system to adapt to recent process dynamics while maintaining interpretability. \newline

\noindent \textbf{Fusion Agent} \newline
The Fusion Agent serves as the central orchestrator, responsible for integrating outputs from all agents to generate the final WiP prediction. Using a ReAct-style prompting mechanism, it combines intermediate predictions and qualitative trend insights through multi-step reasoning. The Fusion Agent performs the following operations:
\begin{enumerate}
    \item Queries all Predictor Agents and collects their individual predictions.
    \item Retrieves trend descriptions from the Decision-Making Assistant Agent.
    \item Compares and contrasts the numerical predictions and trend indicators.
    \item Applies decision rules or ensemble heuristics to resolve discrepancies–e.g., prioritizing WindowedAgent in stable periods and DailyMemoryAgent during rapid shifts.
    \item Delivers the final forecast and, optionally, a rationale detailing how each agent was weighted.
\end{enumerate}

This strategy enables adaptive, interpretable fusion of agent outputs and supports seamless integration of new agents without altering the core architecture.


\subsection{Results}
This subsection reports our framework’s performance on both datasets, comparing agent accuracy and adaptability via visuals and error metrics.  \newline

  
  

\begin{figure}[!t]
  \centering
  \includegraphics[width=\linewidth]{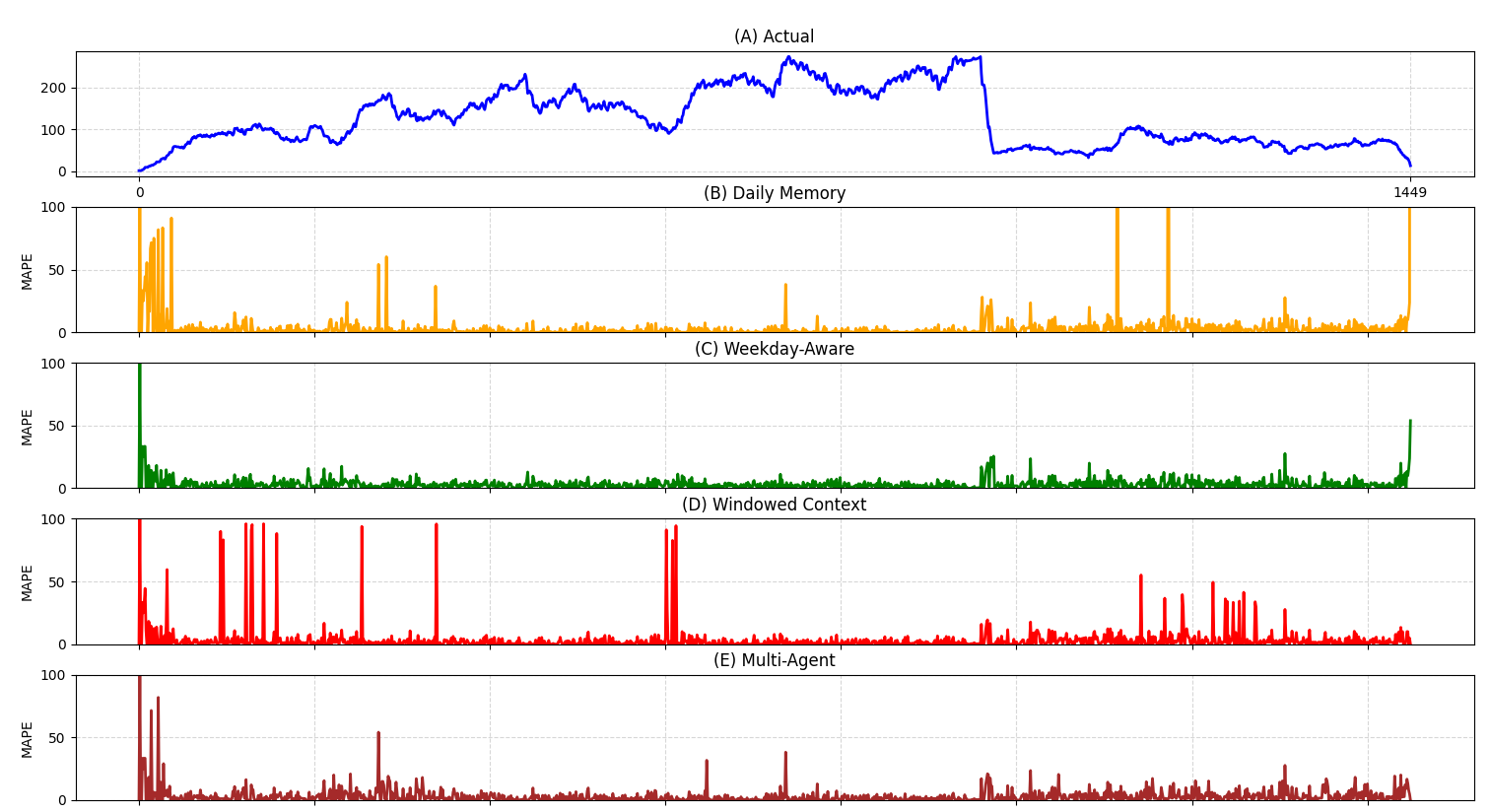}
  \caption{Prediction comparison for the Helpdesk dataset. Top panel shows the actual WiP; bottom panel displays the MAPE error for Daily Memory, Weekday-Aware, Windowed Context, and Multi-Agent models.}
  \label{fig:helpdesk-results}
\end{figure}

\begin{figure}[!t]
  \centering
  \includegraphics[width=\linewidth]{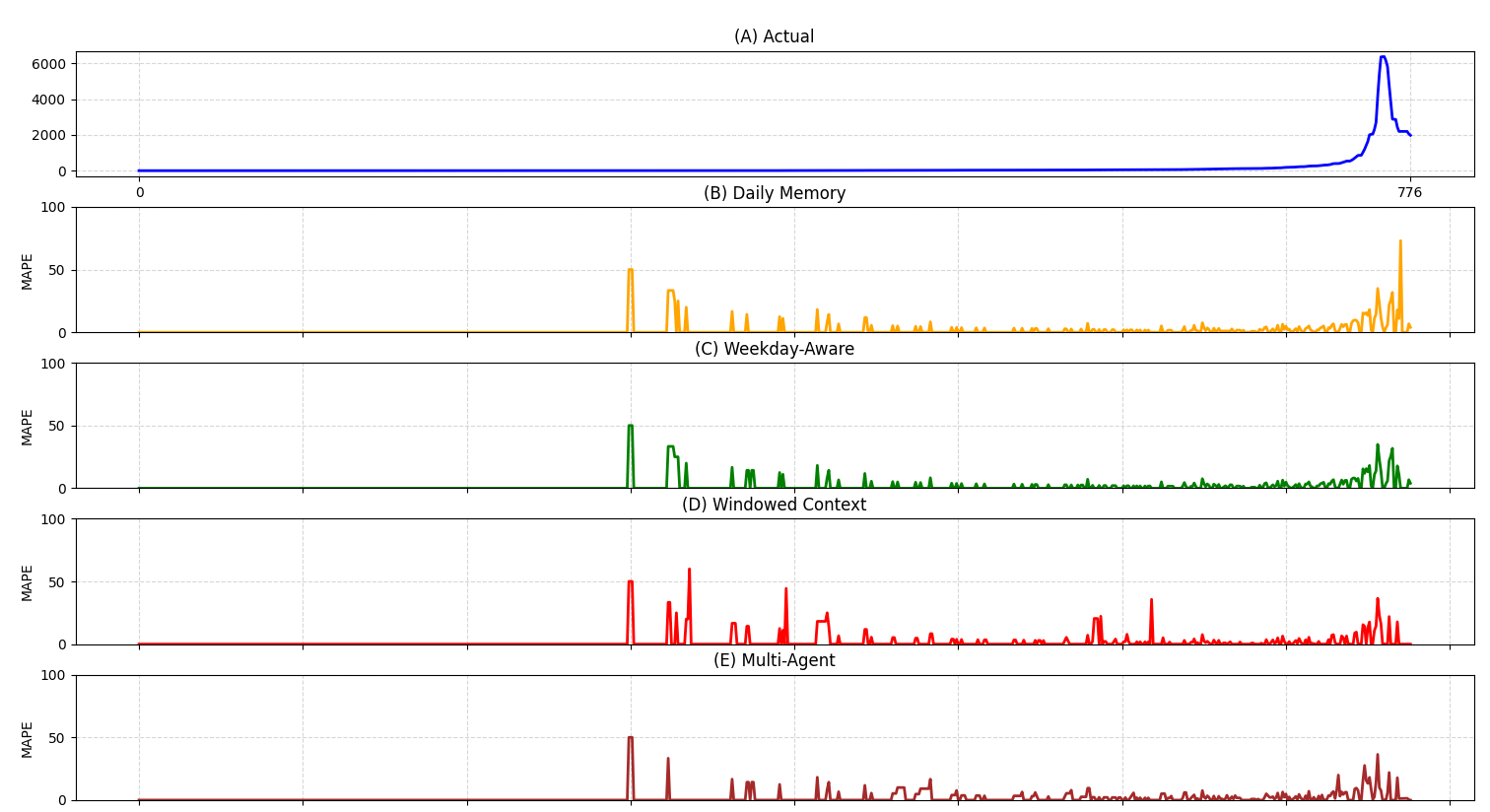}
  \caption{Prediction comparison for the BPIC13 Incidents dataset. Top panel shows the actual WiP; bottom panel displays the MAPE error for Daily Memory, Weekday-Aware, Windowed Context, and Multi-Agent models.} 
  \label{fig:incident-results}
\end{figure}

\noindent\textbf{Visual Comparison of Model Predictions} \newline
Figures \ref{fig:helpdesk-results} and \ref{fig:incident-results} display the actual WiP values and corresponding MAPE trends for each model configuration across the Helpdesk and BPIC13 Incidents datasets. The top panel in each figure shows the actual WiP value, followed by model-specific MAPE plots for Daily Memory, Weekday-Aware, Windowed Context, and Fusion or Multi-Agent.

In the Helpdesk dataset, Daily Memory displays large error spikes during unstable periods, reflecting its limited ability to adapt to rapid changes.  Adding weekday context reduces early variance by capturing cyclical patterns, while the windowed model improves mid-range performance but still suffers from spikes. In contrast, the Multi-Agent model shows the most stable behavior, maintaining a consistently low MAPE throughout the timeline, demonstrating its robustness across volatile periods. In the BPIC13 Incidents dataset, all methods experience difficulty during the sharp workload increase near the end. However, the multi-agent model maintains significantly lower error across this transition, underscoring its strength in generalizing over abrupt process changes by leveraging multiple memory and prediction pathways, outperforming simpler models during regime shifts.

These visual trends reveal a clear performance hierarchy: the Daily Memory model exhibits the highest error spikes, followed by moderate improvements in the Weekday-Aware and Windowed Context models, with the Multi-Agent model achieving the most consistent and lowest MAPE. These observations support the quantitative findings in Table \ref{tb:prediction-results}; simple memory models struggle with dynamics and regime shifts, while the Multi-Agent framework delivers robust and adaptive performance across both datasets, achieving the lowest MAPE and MAE, confirming its superior adaptability in dynamic process environments.
\\

\begin{table}[!b]
\caption{Prediction Results (MAPE and MAE) for Helpdesk and BPIC13 Datasets}
\label{tb:prediction-results}
\centering
\begin{tabular}{|p{0.35\columnwidth}|p{0.15\columnwidth}|p{0.15\columnwidth}|p{0.15\columnwidth}|p{0.15\columnwidth}|}
\hline
\textbf{Model} & \multicolumn{2}{c|}{\textbf{Helpdesk}} & \multicolumn{2}{c|}{\textbf{BPIC13 Incidents}} \\
\cline{2-5}
 & \quad \textbf{MAPE} & \quad \textbf{MAE} & \quad \textbf{MAPE} & \quad \textbf{MAE} \\
\hline
\multicolumn{5}{|l|}{\textbf{Benchmarks}} \\ \hline

TCN\#1 & \quad 2.35 & \quad -- & \quad 1.53 & \quad -- \\ \hline

TCN\#2 & \quad 2.39 & \quad -- & \quad 2.94 & \quad -- \\ \hline

TCN\#3 & \quad 2.45 & \quad -- & \quad 3.14 & \quad -- \\ \hline

LSTM & \quad 34.15 & \quad -- & \quad 2.99 & \quad -- \\ \hline
Persistent Forecast & \quad 2.65 & \quad -- & \quad 0.86 & \quad -- \\ \hline

\hline
\multicolumn{5}{|l|}{\textbf{Proposed Methods}} \\ \hline
Daily Memory Predictor & \quad 4.34 & \quad 2.62 & \quad 1.59 & \quad 15.93 \\ \hline
Weekday-Aware Predictor & \quad 3.13 & \quad 3.29 & \quad 1.55 & \quad 13.55 \\ \hline
Windowed Context Predictor & \quad 3.19 & \quad 3.08 & \quad 1.66 & \quad 9.56 \\ \hline
Multi-Agent Predictor & \quad \textbf{2.91} & \quad \textbf{2.90} & \quad \textbf{1.50} & \quad \textbf{9.45} \\ \hline
\end{tabular}
\end{table}


\noindent\textbf{Quantitative Performance Comparison}\newline 
Table \ref{tb:prediction-results} presents the quantitative evaluation of all models across the Helpdesk and BPIC13 Incidents datasets using MAPE and MAE. The benchmark models-drawn from previous studies-primarily report MAPE and generally fall in the range of 2.3–3.1\% on both datasets, with Persistent Forecast performing best on the BPIC13 Incidents dataset.

Among our proposed methods, the Fusion Predictor consistently outperforms the others, achieving the lowest MAPE and MAE across both datasets. On the BPIC13 Incidents dataset, it reduces the MAPE to 1.50\%, outperforming all benchmarks and baselines. It also achieves the lowest MAE (9.45), indicating stable predictions even during sharp fluctuations. For the Helpdesk dataset, while all our methods yield slightly higher MAPE than the benchmarks, the Fusion model still delivers the best performance within our framework (MAPE = 2.91, MAE = 2.90). Notably, the LSTM model-despite being a deep learning-based approach-performs poorly on the Helpdesk dataset (MAPE = 34.15), likely due to data sparsity or lack of temporal consistency. These results highlight the Fusion Predictor’s ability to generalize across dynamic process regimes, leveraging multiple contextual retrieval strategies and predictive agents to achieve superior adaptability, as evidenced by its consistently low error rates in both stable and volatile conditions.

Overall, these results confirm the benefit of combining multiple contextual retrieval strategies and predictive agents. While simple models such as Daily Memory or Persistent Forecast can achieve reasonable performance under stable conditions, they struggle in volatile regions. In contrast, the Fusion Predictor generalizes better across dynamic process regimes, as supported by both numerical and visual evaluations.

\section{Conclusion and Future Work}
We proposed a novel framework for predictive process monitoring that integrates RAG with a multi-agent architecture to predict WiP. By converting structured event logs into natural language stories and deploying agents focused on distinct temporal views, the framework enhances predictive accuracy and interoperability through semantic reasoning. Key advantages include dynamic, context-aware retrieval via a process memory, extensibility through agent-based modularity, and robust decision fusion that adapts to process trends. Experiments on real-world datasets validate the effectiveness of the approach, highlighting its competitiveness and the benefit of combining symbolic and neural reasoning.

Future directions include incorporating adaptive retrieval mechanisms that optimize memory usage over time, enabling interactive forecasting through user feedback, and extending the architecture to support additional predictive tasks such as cycle time estimation, anomaly detection, and compliance monitoring. Integration with domain-specific knowledge graphs may also enhance the transparency and reliability of the predictions.

\end{document}